\documentclass[reprint, amsmath,amssymb,showpacs,floatfix, aps, prb, twocolumn, superscriptaddress,longbibliography]{revtex4-2}
\usepackage{graphicx}
\usepackage{dcolumn}
\usepackage{bm}
\usepackage{color}
\usepackage{hyperref}
\usepackage[normalem]{ulem}
\hypersetup{colorlinks=true,citecolor=blue,linkcolor=blue, urlcolor=blue}
\newcolumntype{M}[1]{>{\centering\arraybackslash}m{#1}}
\newcolumntype{N}{@{}m{0pt}@{}}
\def\nn{\nonumber}
\def\({\left(}
\def\){\right)}
\def\[{\left[}
\def\]{\right]}
\def\a{\alpha}
\def\b{\beta}
\def\d{\delta}

\def\s{\sigma}

\def\w{\omega}

\def\bfB{\mathbf{B}}
\def\bfQ{\mathbf{Q}}

\def\bfb{\mathbf{b}}
\def\bfe{\mathbf{e}}
\def\bfk{\mathbf{k}}
\def\bfm{\mathbf{m}}
\def\bfn{\mathbf{n}}

\def\bfq{\mathbf{q}}
\def\bfr{\mathbf{r}}
\def\bfR{\mathbf{R}}

\def\bfx{\mathbf{x}}

\def\bfS{\mathbf{S}}

\def\>{\rangle}
\def\<{\langle}

\begin{document}

\title{Topological Hall effect of Skyrmions from First Principles}

\author{Hsiao-Yi Chen}
\affiliation {RIKEN Center for Emergent Matter Science (CEMS), Wako 351-0198, Japan}

\author{Takuya Nomoto}
\affiliation {Department of Physics, Tokyo Metropolitan University, Hachioji, Tokyo 192-0397, Japan}


\author{Max Hirschberger}
\affiliation {RIKEN Center for Emergent Matter Science (CEMS), Wako 351-0198, Japan}
\affiliation{Department of Applied Physics and Quantum-Phase Electronics Center, The University of Tokyo, Bunkyo-ku, Tokyo 113-8656, Japan}

\author{Ryotaro Arita}
\affiliation {RIKEN Center for Emergent Matter Science (CEMS), Wako 351-0198, Japan}
\affiliation {Department of Physics, The University of Tokyo, Bunkyo-ku, Tokyo 113-0033, Japan}

\date{\today}

\begin{abstract}
We formulate a first-principles approach for calculating the topological Hall effect (THE) in magnets with noncollinear nanoscale spin textures. We employ a modeling method to determine the effective magnetic field induced by the spin texture, thereby circumventing the computational challenges associated with superlattice calculations. Based on these results, we construct a Wannier tight-binding Hamiltonian to characterize the electronic states and calculate the Hall conductivity. Applying this approach to the skyrmion material $\rm Gd_2PdSi_3$ shows good agreement with experimental data. Our analysis in momentum space further reveals that the dominant contribution to the THE arises from the crossing points between the folded bands along high-symmetry lines in the Brillouin zone. 
This work advances numerical techniques for simulating general magnetic system, examplified by but not restricted to skyrmion lattice, and its result offering insights into the complex interplay between spin textures and electronic transport.
\end{abstract}
\maketitle
\section{Introduction \label{Sec:Intro}}
\vspace{-10pt}
The Hall effect, the transverse motion of conducting electrons perpendicular to the direction of an imposed electric field, has been a focal point of research in solid state physics for several decades. While the normal Hall effect (NHE), attributed to the Lorentz force resulting from an external magnetic field, has been comprehensively understood since the 19th century \cite{hall1879new} and aligns with observations in nonmagnetic metals, challenges persist in unraveling the transport properties of ferromagnets \cite{nagaosa2010anomalous}. In collinear ferromagnets, the Hall conductivity persists without an external magnetic field, indicating a different underlying mechanism, which is elucidated as a interplay between intrinsic magnetization and spin-orbit coupling (SOC). In the extrinsic mechanism, this anomalous Hall effect (AHE) is attributed to processes such as skew scattering \cite{karplus1954hall,luttinger1958theory} and the side jump process \cite{berger1970side}. In the intrinsic mechanism, the AHE arises from the effective magnetic field, i.e., the Berry curvature of the Bloch states~\cite{sundaram1999wave}.  On the other hand, in recent years, researchers identify a distinct form of the Hall effect, known as the topological Hall effect (THE)\cite{bruno2004topological} induced by the spin topology in chiral magnets \cite{taguchi2001spin,taguchi2001enhancement, tokura2020magnetic}.
\\
\indent
In noncollinear chiral magnets, spatial spin orientations {can} give rise to two-dimensional spin configurations characterized by the topological charge:
\begin{equation}
    n_{Sk}=\frac{1}{4\pi} \int
    d^2 \bfr~\bar{\bfm}(\bfr)\cdot
    \left[
    \frac{\partial \bar{\bfm}(\bfr)}{\partial x}\times 
    \frac{\partial \bar{\bfm}(\bfr)}{\partial y}
    \right]
    \label{Eq:Nsk}
\end{equation}
where $\bar{\bfm}={\bfm}/{|\bfm|}$, and $\bfm(\bfr)$ is the local magnetization. As an integral number, finite $n_{Sk}$ protects the  topological spin configurations from continuous deformation to a trivial state ($n_{Sk}=0$). 
The skyrmion~\cite{bogdanov1989thermodynamically} is such a topological object, 
which is stabilized by various mechanisms depending on the symmetry of {the crystal host}, including the Dzyaloshinskii-Moriya interaction \cite{nagaosa2013topological} in B20-type compounds \cite{muhlbauer2009skyrmion,neubauer2009topological} and the RKKY-exchange interaction \cite{okubo2012multiple} in centrosymmetric Gd-based magnets \cite{kurumaji2019skyrmion,hirschberger2019skyrmion,khanh2020nanometric}.
By interacting with carriers passing by, the skyrmion can induce the THE. From a semi-classical point of view, the adiabatic transport of electron through a skyrmion structure  gains a Berry phase proportional to the topological charge $N$. This "real space" Berry phase contributes to an effective magnetic field for the transport carriers and causes the transverse deflection \cite{ishizuka2018spin,verma2022unified}, which is a unique property of magnets with chiral spin textures and adopted as a crucial quantity to search skyrmions and other topologically nontrivial spin structures.
\\
\indent
While a wide variety of studies on skyrmion systems based on a phenomenological model~\cite{neubauer2009topological} or microscopic models~\cite{ishizuka2018spin,verma2022unified,ye1999berry,ritz2013giant,tatara2002chirality,denisov2018general,onoda2004anomalous,hamamoto2015quantized,matsui2021skyrmion,onoda2004anomalous,ohgushi2000spin,wang2020skyrmion,hamamoto2015quantized} have been performed, first-principles calculations based on density functional theory (DFT) have been a significant challenge due to the huge system size~\cite{koretsune2015control,jia2018first,nomoto2020formation,10.1063/5.0141628}.
In this study, we introduce an $ab~initio$ methodology for computing the electronic structure and the THE in skyrmion materials. Our approach begins with developing an algorithm to compute the self-consistent Kohn-Sham (KS) potential, for which the magnetic part, $B_{\rm KS}$, is induced by the spin textures. To avoid the computationally intensive task of DFT calculation for nano-sized skyrmion, we propose an ansatz that simplifies $B_{\rm KS}$ as a function of neighboring localized spins, which 
enables efficient acquisition of the magnetic potential to construct the KS-equation. Subsequently, we use the KS orbitals and $B_{\rm KS}$ to construct the  Wannier tight-binding (TB) model and formulate a Hamiltonian that can accurately describe the electronic structure in the skyrmion phase. Finally, using the TB Hamiltonian, we can compute the Berry phase, as well as the topological Hall effect (THE) by the Kubo formula \footnote{The real space Berry curvature from the noncoplaner spinors can be converted into the momentum space when we fully diagonalize the Hamiltonian. See Ref.~\cite{verma2022unified}}. 
\\
\indent
In the second part of this paper, we apply the formalism to the case of $\rm Gd_2PdSi_3$. In the skyrmion phase, our computed spin-polarization {$P$} {is in contradiction with} the magnitude $P$ deduced from the {experimental data based on a }phenomenological approach; rather, the calculated THE exhibits good agreement with experimental measurements, when considering a reasonable shift of the chemical potential. This favors a description in the momentum space picture. Furthermore, we conduct an analysis of the THE contribution in momentum space, revealing that the primary contribution arises along a high-symmetry line,  where a gap opens at the intersection among folded bands when the lattice is extended to a superlattice by the presence of the skyrmion. 
Overall, our work introduces a widely applicable approach for accessing the electronic structure of the skyrmion state and contributes insights into the numerical methods employed to elucidate the transport properties of chiral magnets.
\\
\indent
The structure of this paper is as follows. In Section~\ref{Sec:Theory}, we provide a derivation of the theoretical formalism employed in this study, which includes the modeling method utilized to obtain the effective magnetic field and the extension of the TB Hamiltonian to incorporate the skyrmion effect. In Section~\ref{Sect:Gd213}, we present and discuss the numerical results obtained by applying the developed formalism to the $\rm Gd_2PdSi_3$ system. We summarize the discussion in Sec.~\ref{sect:conclusions}.  In the appendices, we gather the technical details for algorithm implementations and the numerical simulations, and we also present the discussion on how to chose the appropriate Fermi energy in simulation to compare with the experimental measurement. 

\section{Theoretical methods \label{Sec:Theory}}
\vspace{-10pt}

The development of DFT with the noncollinear exchange-correlation (EX) potential~\cite{kubler1988density} has significantly advanced the study of various problems in magnets such as noncollinear magnetism~\cite{hobbs2000fully}, relativistic effects~\cite{richter1998band}, and adiabatic spin dynamics~\cite{halilov1998adiabatic}.
It writes the one-particle KS equation as
\begin{align}
    \sum_{s'}\left\{\left[\frac{-\nabla^2}{2}+V_{\rm KS}(\bfx)\right]\delta_{ss'}+\left[{\boldsymbol{ \sigma}}_{ss'}\cdot \bfB_{\rm KS}(\bfx)\right]\right\}\Psi_{s'}(\bfx)~~~&\nn\\
    =E\Psi_s(\bfx)&
    \label{Eq:KS-noncol}
\end{align}
where $V_{\rm KS}$ is the local EX scalar field and $\bfB_{\rm KS}$ is the local EX magnetic (EX B) field, which are functionals of local charge densities and local magnetism, and $\boldsymbol{\sigma}_{ss'}$ are the Pauli matrices with $s$ denoting the spin component. In practical calculations, $V_{\rm KS}$ and $\bfB_{\rm KS}$ are determined by solving the KS equation self-consistently: starting with an initial guess for the density to generate the potentials, solving Eq.~(\ref{Eq:KS-noncol}) to obtain new densities and update the potentials, and repeating the processes until convergence is reached. Therefore, a DFT calculation is generally heavy, and the computation can grow along with the size of the target system. Consequently, in solid state system, DFT calculations usually adopt translation symmetry to reduce memory requirements by restricting calculations within an unit cell under the formalism of the Bloch theorem. 
\\
\indent
However, noncollinear nano-scale spin textures, exemplified by skyrmions, typically disrupt the crystal translation symmetry, spanning a size that covers $\mathcal{O}(10^2)$ unit cells or more. As a result, the application of the Bloch theorem fails or becomes inefficient, rendering a direct DFT calculation for skyrmionic systems impractical. Thus, current studies can only rely on empirical modeling based on the double exchange Hamiltonian~\cite{matsui2021skyrmion},
\begin{equation}
    H=\sum_{\<r,r'\>} t c^\dagger_rc_{r'}-\frac{J_H}{2}\sum_{r}\bfS_r\cdot\left[c^\dagger_r\boldsymbol{\s} c_r\right],
    \label{Eq:Db-ex}
\end{equation}
where $t$ is the tight-binding hopping amplitude, $\bfS_r$ is the local spin at $r$-site with the coupling strength $J_H$. Although this model can now reveal several physical properties of non-planer spin textures, the undetermined parameters set aside a quantitative prediction in real systems. 
\\
\indent
To overcome this challenge, in Sect.~\ref{Subsec:Model}, we depict an approach to incorporate DFT with the concept of the double exchange model by a modeling procedure. The resultant model function can generate EX potentials induced by spin textures and enables us to obtain the KS equation without going through the self-consistent field (SCF) calculation. 
Furthermore, in order to {calculate the THE from the Kubo formula}, interpolation using a Wannier TB model is necessary. Therefore, in Sect.~\ref{Subsect:wannier}, we present a short review about the approach. Last, in Sect.~\ref{Subsect:sk_tb}, we carry out a mapping procedure which can project the TB model built from a single unit cell system to the one for super cell system. This shows how to introduce the skyrmion potential into the TB model in describing the skyrmion system.

\subsection{Modeling exchange-correlation potential in noncollinear DFT \label{Subsec:Model}}
\begin{figure}[t]
    \centering
    \includegraphics[scale=0.4]{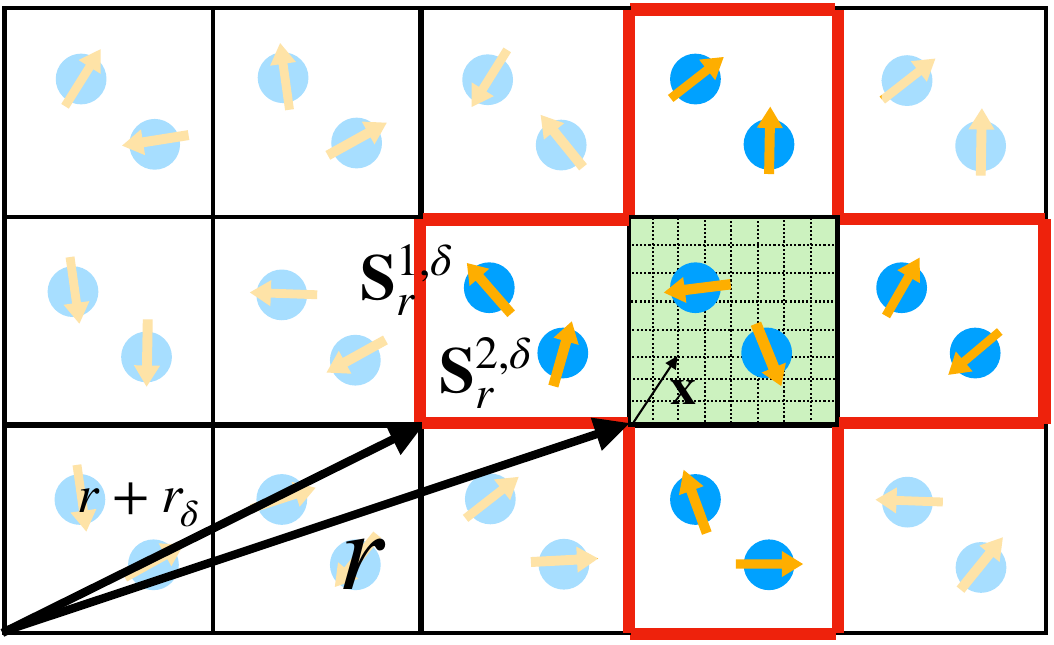}
    \caption{Illustration of Eq.~(\ref{Eq:Bex_S}) in  an example where one unit cell contains two magnetic atoms. We use the blue circles to denote the atoms and the yellow arrows to denote the localized spins. In the unit cell at $r$ (greened region), the EX $B$-field is determined by spinors of all neighboring cell ${\bfS}^{a,\delta}_r$ (enclosed by the red line) via the function ${\bfB}_S(\bfx)$.}
    \label{Fig:coordinate}
\end{figure}

\vspace{-10pt}
In this section, we present a modeling approach to describe the EX potentials of {magnets with chiral spin textures}. We first note that, among recent experiment researches, skyrmion lattices are mainly realized in magnetic compounds with atoms of half-filled $d$ or $f$ orbitals. In these kinds of materials, the magnetism is determined by the spatially localized $d,f$ electrons which are also energetically separated from the conducting electrons \cite{you2022gadolinium}. Therefore, we can decouple the localized electron density and magnetization from other electrons and decompose the EX potentials as:
\begin{align}
   & V_{\rm KS}[n(\bfx)]=V^{\rm loc}_{\rm KS}[n^{\rm loc}(\bfx)]+V^{\rm other}_{\rm KS}[n^{\rm other}(\bfx)]\nn\\
   & {\bfB}_{\rm KS}[\bfm(\bfx)]={\bfB}^{\rm loc}_{\rm KS}[\bfm^{\rm loc}(\bfx)]
   \label{Eq:VB}
\end{align}
where we ignore the magnetization from electrons other than localized magnetic electrons. Besides, based on the numerical result shown in Appendix \ref{Append:fitting}, we find that the scalar potential $V_{\rm KS}$ is almost independent of the magnetization $\bfm(x)$. Therefore, in this work we rewrite the KS Hamiltonian as
\begin{align}
    \left[\frac{-\nabla^2}{2}+V_{\rm KS}(\bfx)\right]\Psi_s(\bfx)+\sum_{s'}\left\{\boldsymbol{\sigma}\cdot {\bfB}^{\rm loc}_{\rm KS}[\bfm^{\rm loc}(\bfx)]\right\}_{ss'}\Psi_{s'}(\bfx)&\nn\\
    =E\Psi_s(\bfx).&
    \label{Eq:KS-noncol-decomp}
\end{align}
By comparing Eq.~(\ref{Eq:KS-noncol-decomp}) with Eq.~(\ref{Eq:Db-ex}), we can observe the similarity between the KS-equation and the double exchange model; the first term is a isotropic term, and the second term is determined by the localized magnetization. Consequently, learning from the double exchange model, we propose that the self-consistent potential ${\bfB}_{\rm KS}[\bfm (\bfx)]$ within a certain unit cell at site $r$ (${\bfB}^r_{\rm KS}$) can be determined by localized spins via a model function:
\begin{eqnarray}
    &&{\bfB}^r_{\rm KS}[\bfm (\bfx)]={\bfB}_{S}[\{\bfS\}_r](\bfx)\nn\\
    &&\{\bfS\}_r=\left(\bfS^{a=1,\delta=0}_r,\bfS^{a=2,\delta=0}_r,\cdots, \bfS^{a=n_a,\delta=n}_r\right)
    \label{Eq:Bex_S}
\end{eqnarray}
where $n_a$ is the number of magnetic atoms in a unit cell and $n$ is the number of neighboring cells. Meanwhile, $\{\bfS\}_r$ denotes the spin set with $\bfS^{a,\delta}_r$ being the localized spin of the $a$-th magnetic atom in the $\delta$-th neighbor cell of the cell at $r$, defined as:
\begin{equation}
    \bfS^{a,\delta}_r=\int_{\bfx\in R_{a,r+r_\delta}}  \bfm(\bfx)~d\bfx
    \label{Eq:S-m}
\end{equation}
where $R_{a,r+r_\delta}$ is a point enclosed by a sphere with radius $R$ centered at the $a$-th atom in the unit cell at $r+r_\delta$ \footnote{In this work, the radius $R_{a,r+r_\delta}$ is determined as the minimum distance between the $a$-th atom and the neighboring atom divided by two and multiplied by 1.2 as implemented in {\sc quantum espresso} package.} (See Fig.~\ref{Fig:coordinate}). Here, $\delta=0$ denotes the home cell, i.e. the unit cell at $r$ itself, and $r_{\delta=0}=0$. It should be noted that the ansatz function ${\bfB}_S$ is the same for each unit cell such that it respects all the symmetries of the crystal structure, including translation and rotational symmetry. Using Eq.~(\ref{Eq:Bex_S}), without a self-consistent calculation, we can  generate a self-consistent EX $B$-potential from a given spin orientation $\{\bfS\}$ and solve Eq.~(\ref{Eq:KS-noncol-decomp}) to get the band structure and compute corresponding physical quantities.
\\
\indent
The ansatz for Eq.~(\ref{Eq:Bex_S}) is arbitrary, so long as the function can reproduce the DFT result up to required accuracy. For a crystal which contains one magnetic atom in an unit cell, the simplest ansatz is:
\begin{equation}
    {\bfB}_{S}[\{\bfS\}_r](\bfx)=\frac{J_H}{2}\bfS_r^{a=1,\delta=0}
\end{equation}
For a unit cell at site $r$ with more than one magnetic atom, we use the following ansatz in the present study:
\begin{equation}
    {\bfB}_{S}[\{\bfS\}_r](\bfx)=\sum_{a,\delta} \stackrel{\leftrightarrow}{\bf J}_{a,\delta}(\bfx)\cdot \bfS^{a,\delta}_r
    \label{Eq:B_JS}
\end{equation}
where 
$\stackrel{\leftrightarrow}{\bf J}_{a,\delta}(\bfx)$ is a tensor function of the real space grid in a unit cell, which depends on the associated atom index $a$ and neighboring cell index $\delta$.
The details of $\stackrel{\leftrightarrow}{\bf J}_{a,\delta}(\bfx)$ can be refined by sophisticated theories but depends on crystal parameters and varies among chemical compounds. Therefore, we leave it {adjustable} and determine its form numerically. In practical computations, we first conduct several self-consistent DFT calculations using randomly generated $\{\bfS\}_r$ sets, and fit the ansatz function Eq.~(\ref{Eq:B_JS}) to obtain the coefficients $\stackrel{\leftrightarrow}{\bf J}_{a,\delta}(\bfx)$ on a real space grid.
\\
\subsection{Wannierization, Berry connection and Hall conductance \label{Subsect:wannier}}
\vspace{-10pt}
Computing the Berry curvature from first principles requires an extremely high computational cost because the Berry curvature varies drastically in \textbf{k}-space. Besides, additional effort must be taken to make sure that the gauge of the wave functions varies smoothly for the covariant derivative \cite{fang2003anomalous,Yao2004first}. To overcome these problems, the interpolation method using the localized Wannier function \cite{wang2006ab} has provided a low-cost approach. In this section, we briefly review the formalism and then introduce our approach to extend it to a superlattice system to include the skyrmion potential in the next section. We define the Bloch function in the Wannier gauge as a linear combination of localized Wannier functions:
\begin{equation}
    |u^{(\rm W)}_{n\bfk}\>=\sum_{\bfR}e^{-i\bfk(\hat{x}-\bfR)}|\bfR n\>
    \label{Eq:w2b}
\end{equation}
where $\bfk$ is in units of the reciprocal vectors, taking values from $0.0$ to $1.0$, $\bfR$ is an integer describing the location of the unit cell, $\hat{x}$ is the position operator, and $n$ is the index of the Wannier function. 
We can write the Hamiltonian in \textbf{k}-space as a Fourier transform of the Wannier TB model,
\begin{equation}
    \<u^{(\rm W)}_{n\bfk}|\hat{H}(\bfk) |u^{(\rm W)}_{m\bfk}\>
    =\sum_{\bfR}e^{i\bfk \cdot \bfR}\<{\bf 0}n|\hat{H}|\bfR m\>
    \label{Eq:HR}
\end{equation}
where $\<{\bf 0}n|\hat{H}|\bfR m\>$ is the hopping amplitude between the $m$-th Wannier function in the $\bfR$-cell and the $n$-th Wannier function in the $\bf 0$-cell. By diagonalizing Eq.~(\ref{Eq:HR})
\begin{equation}
    U^\dagger_{mn_1\bfk }H_{n_1m_1}(\bfk)U_{m_1n\bfk}=\delta_{mn}E_{n\bfk},
\end{equation}
we obtain the band structure {of the Wannier model}, which should be consistent with the DFT result.
\\
\indent
The momentum derivative of the eigenfunction can be correspondingly written as:
\begin{equation}
    |\partial_\a u^{(\rm H)}_{n\bfk}\>=\sum_{m}|\partial_\a u^{(\rm W)}_{m\bfk}\>U_{mn\bfk}+
    \sum_{m}| u^{(\rm H)}_{m\bfk}\>D^{(\rm H)}_{mn\bfk.\a}
\end{equation}
where we simplify the notation by $\partial/\partial {k^\a}=\partial_{\a}$. $W$ and $H$ superscripts indicate the Wannier and Hamiltonian gauges \cite{wang2006ab}, respectively, which differ by a unitary transformation, and
\begin{equation}
    D^{(\rm H)}_{mn\bfk.\a}=(U^\dagger\partial_\a U)_{nm}=
    \left\{
    \begin{matrix}
    \frac{(U^\dagger \partial_\a H^{(\rm W)} U)_{nm}}{E_{m\bfk}-E_{n\bfk}}~~&{\rm if}~~n\neq m\\
    0~~&{\rm if}~~n= m
    \end{matrix}
    \right.
    \label{Eq:Dnm}
\end{equation}
Consequently, the generalized Berry connection can be obtained as:
\begin{equation}
    A_{nm\bfk,\a}=i\<u^{(\rm H)}_{n\bfk}|\partial_\a u^{(\rm H)}_{m\bfk}\>=(U^\dagger A^{(\rm W)}_{\a} U)_{nm}+iD^{(\rm H)}_{mn\bfk,\a}
    \label{Eq:Anm}
\end{equation}
where $A_{nm\bfk,\a}^{(\rm W)}=i\<u^{(\rm W)}_{n\bfk}|\partial_\a u^{(\rm W)}_{m\bfk}\>$. The quantities in Eqs.~(\ref{Eq:Dnm}) and (\ref{Eq:Anm}) can be calculated by taking derivatives of Eqs.~(\ref{Eq:w2b}) and (\ref{Eq:HR}):
\begin{align}
     &\<u^{(\rm W)}_{n\bfk}|\partial_\a \hat{H}(\bfk) |u^{(\rm W)}_{m\bfk}\>
    =\sum_{\bfR} e^{i\bfk \cdot \bfR}i\bfR_\a\<{\bf 0}n|\hat{H}|\bfR m\>
    \label{Eq:HRR}
    \\
    &A_{nm\bfk,\a}^{(\rm W)}=\sum_{\bfR} e^{i\bfk \cdot \bfR}\<{\bf 0}n|\hat{x}_\a|\bfR m\>.
    \label{Eq:ARR}
\end{align}
Solving Eqs.~(\ref{Eq:HR}), (\ref{Eq:Anm}), (\ref{Eq:HRR}), and (\ref{Eq:ARR}), we can compute the Hall conductivity using the Kubo formula \cite{blount1962formalisms}:
\begin{align}
    &\sigma_{\a\b} (\omega)=\frac{ie^2}{\hbar N_k\Omega_c}\sum_{\bfk n, m} (f_{m,\bfk}-f_{n,\bfk})\nn\\
    &~~~~~\times\frac{E_{m\bfk}-E_{n\bfk}}{E_{m\bfk}-E_{n\bfk}-(\hbar\w+i\eta)}A_{nm\bfk,\a}A_{mn\bfk,\b}
    \label{Eq:Kubo}
\end{align}
where $E_{n \bfk}$ is the eigenenergy, and $f_{n,\bfk}$ is the occupation number following the Fermi-Dirac distribution.
For the static limit, we will simply use Eq.~(\ref{Eq:Kubo}) and take $\w \rightarrow 0$.

\subsection{Tight-binding model on a superlattice with skyrmionic magnetic potential \label{Subsect:sk_tb}}

\begin{figure}[t]
    \centering   \includegraphics[scale=0.185]{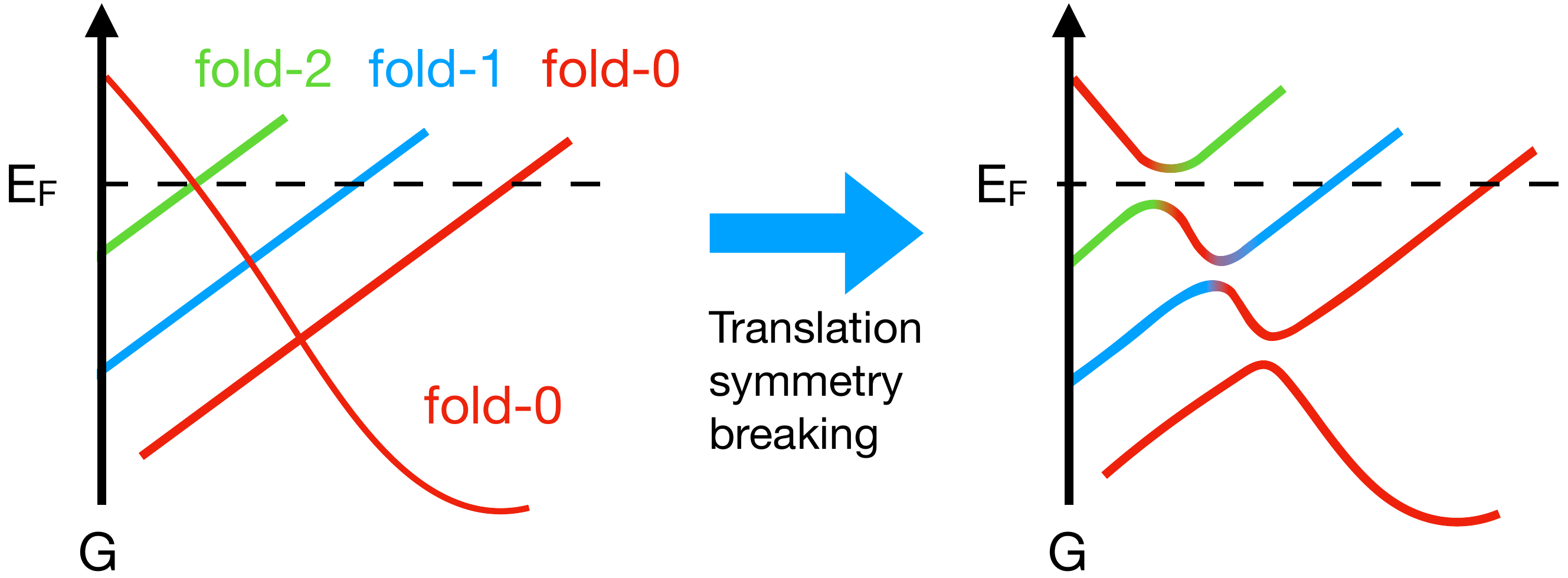}
    \caption{Illustration of band gap-opening due to the skyrmion potential. Band crossings protected by the translation symmetry are lifted such that the manifolds of different folded states (distinguished by colors) are hybridized. A large Berry curvature is expected in the gap region, which contributes to the Hall conductivity when the Fermi-level, $E_{\rm F}$, aligns with the gap.}
    \label{Fig:fold_in}
\end{figure}

Equations~(\ref{Eq:w2b}-\ref{Eq:ARR}) set up the scheme to compute the Berry connections in the primitive cell. However, when a skyrmion lattice is formed, the translation symmetry $\bfx\rightarrow \bfx+\bfR_i$ is broken, and the TB model needs to be modified. In the following, we restrict the discussion to the case that the emergent skyrmion lattice is commensurate with the original primitive cell, such that the periodicity is given by $\bfx\rightarrow \bfx+ N_i\bfR_i$ where $N_i=(N_1,N_2,N_3)$ are the dimensions of the skyrmion along the three crystal axes. In this $N_1\times N_2\times N_3$ supercell, the BZ of the original primitive crystal (pBZ, for the nonmagnetic NM state) will be folded and the \textbf{k}-points are identified by 
\begin{equation}
   \bfk_{\rm p}=\bfk_{\rm s} +n_1\hat{\bfb}_1+n_2\hat{\bfb}_2
+n_3\hat{\bfb}_3\rightarrow \bfk_{\rm s}
\label{Eq:fold-in}
\end{equation}
where $\bfk_{\rm p}$ is a point in the pBZ, $\bfk_{\rm s}$ is a point in the BZ of the supercell (sBZ) \footnote{In the unit of the primitive reciprocal lattice, $\bfk_{\rm s}$ takes value from $(0,0,0)$ to $(1/N_1,1/N_2,1/N_3)$}, $\hat{\bfb}_i=\mathbf{G}_i/{N_i}$
is a reciprocal lattice vector {in the sBZ}, and $n_i= 1\sim N_i$ for each direction. This mapping identifies all the points $\bfk_{\rm p}$ connected by $\hat{\bfb}_i$'s to $\bfk_{\rm s}$ in the sBZ, as well as the Bloch wave functions. To keep track of these states, we introduce a generalized band index (m,$\mathbf{n}$) such that for $\bfk_{\rm p}\in {\rm pBZ}$, we can write the Bloch wave in the sBZ as:
\begin{align}
    &\psi_{m \bfk_{\rm p}}(\bfx)=\psi_{m \bfk_{\rm s}+\bfn\hat{\bfb}}(\bfx)=e^{i(\bfk_s+\bfn\hat{\bfb})\bfx }u^{\rm p}_{m\bfk_{\rm s}+\bfn\hat{\bfb}}(\bfx)
    \nn
    \\
    &=e^{i\bfk_s\bfx }\left[e^{i\bfn\hat{\bfb}\bfx }u^{\rm p}_{m\bfk_{\rm s}+\bfn\hat{\bfb}}(\bfx)\right]\equiv e^{i\bfk_s\bfx } u^{\rm s}_{(m,\bfn),\bfk_{\rm s}}(\bfx).
    \label{Eq:bloch_in_super}
\end{align}
where we use the shorthand $\bfn\hat{\bfb}=\sum_in_i \hat{\bfb}_i$, while $u^{\rm p}$ and $u^{\rm s}$ are the periodic part of the wave function in the primitive lattice and in the superlattice, respectively.
\\
\indent
On the other hand, due to the extension of the cell, we need to redefine the Wannier functions in a unit cell of  the superlattice (i.e. a single supercell). We split the notation of {an atom's position} $\bfR_{\rm p}$ in the primitive crystal by:
\begin{equation}
    \bfR_{\rm p}\rightarrow \bfR_{\rm in}+\bfR_{\rm s}.
    \label{Eq:RpRs}
\end{equation}
$\bfR_{\rm in}$ is the internal translation within a supercell, and $\bfR_{\rm s}$ denotes the position of the supercell. In units of the primitive lattice constant, for each {of the three directions $i$}, {$R_{{\rm in},i}=R_{{\rm p},i}\,\,\mathrm{mod}\, N_i$} takes integer values and $R_{{\rm s},i}=(R_{{\rm p},i}-R_{{\rm in},i})$ takes integer multiples of $N_i$. 
Therefore, we can extend the Wannier orbital index using Eq.~(\ref{Eq:RpRs}) and write the Wannier function as:
\begin{equation}
    |n, \bfR_{\rm p}\>\rightarrow |n,\bfR_{\rm in},\bfR_{\rm s}\>
    \label{Eq:wannier_in_super}
\end{equation}
where $(n,\bfR_{{\rm in}})$ form a collective index to denote the Wannier function in the superlattice.
\\
\indent
Equations (\ref{Eq:bloch_in_super}) and (\ref{Eq:wannier_in_super}) set up the notation for us to extend the Wannier formalism of Ref.~\cite{wang2006ab} to a supercell. We first note that the Bloch functions in the Wannier gauge [Eq.~(\ref{Eq:w2b})] become, from Eq.~(\ref{Eq:bloch_in_super}): 
\begin{align}
    &|\hat{u}^{(\rm W)}_{(m,\bfn), \bfk}\>=e^{i\bfn \hat{\bfb}\cdot\bfx}
    \sum_{\bfR_{\rm p}}e^{-i(\bfk+\bfn\hat{\bfb})(\bfx-\bfR_{\rm p})}|m,\bfR_{\rm p}\>\nn\\
    &=
    e^{i\bfn \hat{\bfb}\bfx}
    \sum_{\bfR_{\rm s},\bfR_{\rm in}}e^{-i(\bfk+\bfn\hat{\bfb})(\bfx-\bfR_{\rm s}-\bfR_{\rm in})}|m,\bfR_{\rm in},\bfR_{\rm s}\>\nn\\
    &=
    \sum_{\bfR_{\rm s},\bfR_{\rm in}} e^{-i\bfk(\bfx-\bfR_{\rm s})}\left[e^{i(\bfk+\bfn\hat{\bfb})\bfR_{\rm in}}\right]|m,\bfR_{\rm in},\bfR_{\rm s}\>
\end{align}
This is the gauge to fix as we extend the primitive cell to a supercell. 
\begin{widetext}
Furthermore, for each $\bfk$ in the sBZ, we can write the Hamiltonian with $\mathbf{R}_1, \mathbf{R}_2\in\{\mathbf{R}_{in}\}$:
\begin{align}   
    &H(\bfk)=\<\hat{u}^{(\rm W)}_{(m_1,\bfn_1)\bfk}|\hat{H}|\hat{u}^{(\rm W)}_{(m_2,\bfn_2)\bfk}\>=
    \sum_{\bfR_{s},\bfR_{1},\bfR_{2}}
    e^{i\bfk\cdot\bfR_{s}}
    e^{i\bfk(\bfR_{2}-\bfR_{1})}
    e^{i(\bfn_2\hat{\bfb}\cdot\bfR_{2}-\bfn_1\hat{\bfb}\cdot\bfR_{1})}
    \<m_1,\bfR_{1},{\bf 0}|\hat{H}|m_2,\bfR_{2},\bfR_{\rm s}\>
    \label{Eq:super-Hq}
\end{align}
Correspondingly, in order to compute the Berry connection, we extend Eq.~(\ref{Eq:HRR}) and Eq.~(\ref{Eq:ARR}):
\begin{align}
    \<\hat{u}^{(\rm W)}_{(m_1,\bfn_1)\bfk}|
    \partial_\a\hat{H}|\hat{u}^{(\rm W)}_{(m_2,\bfn_2)\bfk}\>=
    \sum_{\bfR_{\rm s},\bfR_{1}\bfR_2}
    i(\bfR_{\rm s}+\bfR_{2}-\bfR_{1})_\a~
    e^{i\bfk\bfR_{\rm s}}
    e^{i\bfk(\bfR_2-\bfR_1)}
    e^{i(\bfn_2\hat{\bfb}\bfR_2-\bfn_1\hat{\bfb}\bfR_1)}
    \<m_1,\bfR_1,{\bf 0}|\hat{H}|m_2,\bfR_2,\bfR_{\rm s}\>
    \label{Eq:super-delHq}
\end{align}
and
\begin{align}
    \<\hat{u}^{(\rm W)}_{(m_1,\bfn_1), \bfk}|\partial_\a|\hat{u}^{(\rm W)}_{(m_2,\bfn_2), \bfk}\>=
    \sum_{\bfR_{\rm s},\bfR_{1}\bfR_2}
    e^{i\bfk\bfR_{\rm s}}
    e^{i\bfk(\bfR_2-\bfR_1)}
    e^{i(\bfn_2\hat{\bfb}\bfR_2-\bfn_1\hat{\bfb}\bfR_1)}
    \<m_1,{\bf 0},{\bf 0}| \hat{x}_\a
    |m_2,\bfR_2-\bfR_1,\bfR_{\rm s}\>
    \label{Eq:super-AAq}
\end{align}
\end{widetext}
It is worth noting that Eqs.~(\ref{Eq:super-Hq}-\ref{Eq:super-AAq}) contain no more information than the unit cell construction, and describe the same physics as Eqs.~(\ref{Eq:HR},\ref{Eq:HRR}-\ref{Eq:ARR}), but with a larger cell. Rather, the skyrmion effect can be acquired once we replace the magnetic potential in $\hat{H}$ with the EX B-field generated by the skyrmion. In practical implementation, we set up the primitive TB Hamiltonian in a non-magnetic (NM) phase and add the EX $B$-field of Eq.~(\ref{Eq:B_JS}), which is rewritten as:
\begin{equation}
    {\bfB}^{\rm super}_S(\bfx_{\rm s})=\sum_{\bfR_{\rm in}}\sum_{a,\delta} \stackrel{\leftrightarrow}{\bf J}_{a,\delta}(\bfx_{\rm p})\cdot \bfS^{a,\delta}_{\bfR_{\rm s}+\bfR_{\rm in}}
    \label{Eq:B_JS2}
\end{equation}
where $\bfS^{a,\delta}_{\bfR}$ is the spin texture deduced from experimental data, $\bfx_{\rm s}$ is the real space grid in superlattice, and $\bfx_{\rm p}$ is the the real space grid in the primitive cell, which are related by:
\begin{equation}
    \bfx_{\rm s}=\bfx_{\rm p}+\bfR_{\rm in}+\bfR_{\rm s}.
\end{equation}
Therefore, the hopping amplitude  between two Wannier functions, induced by the skyrmion lattice, becomes:
\begin{align}
    &\<m_1,\bfR_1,{\bf 0}|\sum_{\bfx_{\rm s}}\boldsymbol{\sigma}\cdot{\bfB}^{\rm super}_S(\bfx_{\rm s})|m_2,\bfR_2,\bfR_{\rm s}\>\nn\\
    &=
\sum_{\bfx_{\rm p},\bfR'_s,\bfR_{\rm in}}\boldsymbol{\sigma}\cdot\nn\\
&~~~
 \<m_1,\bfR_1,{\bf 0}|{\bfB}^{\rm super}_S(\bfx_{\rm p}+\bfR_{\rm in}+\bfR'_{\rm s})
|m_2,\bfR_2,\bfR_{\rm s}\>.
\label{Eq:mBm}
\end{align}
Although the the summation runs over  the entire real space grid, we can use the localized property of Wannier functions and truncate the summation. In this work, we restrict the EX $B$-field effect to act only on the second-nearest neighbor site, i.e. for each direction $\a$:
\begin{align}
    &|(\bfR_2+\bfR_{\rm s}-\bfR_{\rm in}-\bfR'_{\rm s})_\a|\leq 1\nn\\
    &|(\bfR_1+{\bf 0}-\bfR_{\rm in}-\bfR'_{\rm s})_\a|\leq 1
\end{align}
This truncation operation is one of the most important benefits of computing the EX $B$-field effect in the Wannier basis. If we use the Bloch wave functions, the integration must be done in the full real space of the supercell, which increases the computation cost proportional to the size of skyrmion. On the other hand, the computation cost depends only on the convergence of truncation when the Wannier basis is used.
\begin{figure}[t]
    \centering
    \includegraphics[scale=0.28]{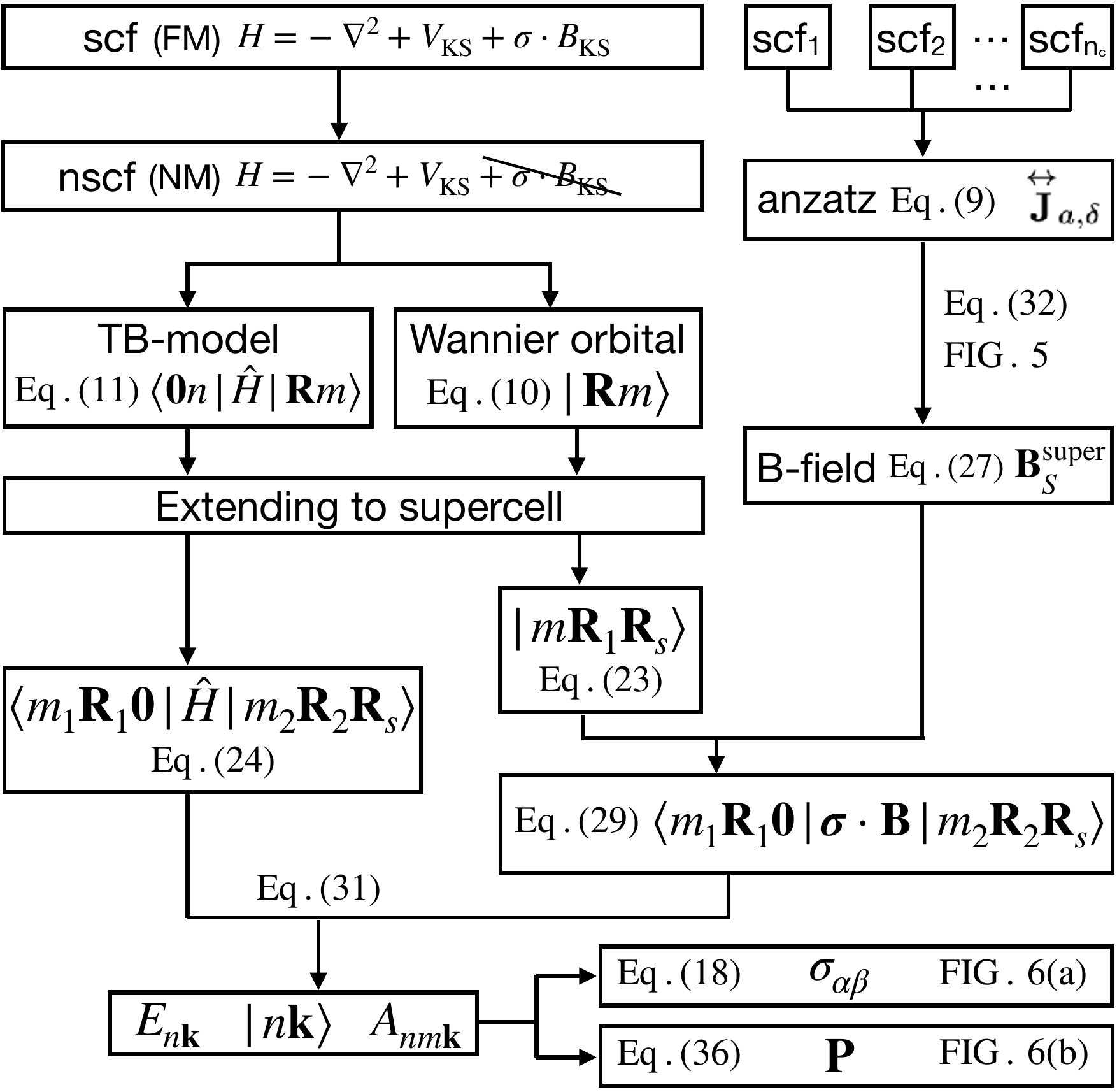}
    \caption{Workflow of the algorithm adopted in this work. We set up the Wannier tight binding (TB) model in the nonmagnetic (NM) phase, which is further extended to a superlattice containing skyrmions. Meanwhile, we conduct $n_c$ self-consistent field (scf) calculations of random spin configurations in a supercell of intermediate size, to fit the ansatz functions which can give us the magnetic EX $B$-field of the skyrmion. The last step combines the TB model with the magnetic field and reproduces the electronic structure of the skyrmion lattice, from which we deduce the electronic bands and the Hall conductivity which can be compared with experiment. }
    \label{Fig:workflow}
\end{figure}
\\
\indent
Overall, Eq.~(\ref{Eq:mBm}) gives us the last piece to compute the Wannier TB model under the skyrmionic effect. By replacing:
\begin{align}
    &\< m_1,\bfR_1,{\bf 0}|\hat{H}|m_2,\bfR_2,\bfR_{\rm s}\>\nn\\
    &~~\rightarrow
    \< m_1,\bfR_1,{\bf 0}|\hat{H}+\boldsymbol{\sigma}\cdot \bfB^{\rm super}_S|m_2,\bfR_2,\bfR_{\rm s}\>
    \label{Eq:HB}
\end{align}
in Eqs.~(\ref{Eq:super-Hq}) and (\ref{Eq:super-delHq}), we can obtain the Hamiltonian of the skyrmion lattice and evaluate the Berry connection as well as the Hall conductivity. As a summary, we visualize the full workflow adopted in this work in Fig.~\ref{Fig:workflow} for a comprehensive understanding. 
\\
\indent
Before closing this section, we compare the Hall effect induced by the EX $B$-field in Eq.~(\ref{Eq:HB}) with the AHE in ferromagnets. It is well-known that the conventional AHE in ferromagnets is enhanced by the SOC which opens a gap at the band nodal point, at which small energy transition in the denominator of Eq.~(\ref{Eq:Kubo}) results in a significant contribution to the Hall conductivity. In contrast, the EX $B$-field in Eq.~(\ref{Eq:HB}) 
hybridizes folded states, opens the energy gap, and induces the Hall effect (See Fig.~\ref{Fig:fold_in}). Besides, in the gap regions,  different folded manifolds are connected to each other such that the Bloch wave function changes drastically and contributes to a large $\<u_{n\bfk}|\partial_k u_{m\bfk}\>$ which greatly enhances the Berry curvature and Hall conductivity.

\section{Topological Hall conductance in $\bf Gd_2PdSi_3$ \label{Sect:Gd213}}
\vspace{-10pt}
Quasi-layered $\rm Gd_2PdSi_3$ exhibits magnetic phase transitions characterized by the re-orientation of localized 4$f$ spins of the Gd atom. A recent experiment found that when applying external magnetic field to $\rm Gd_2PdSi_3$, kink-like changes in the magnetization was observed, which indicate two first order phase transitions of the spin structure \cite{kurumaji2019skyrmion,hirschberger2020topological,hirschberger_phasediagram_2020b}. By comparing the measured transverse conductivity with the normal and anomalous Hall conductivity calculated in a phenomenological model, a large deviation appears in the field-induced second phase. This is attributed to the topological Hall effect originating from the formation of a non-trivial topological spin texture. Using x-ray scattering measurements, a spin-modulation of $\sim 2.5\,$nm along three directions of a hexagonal lattice is identified, which suggests the existence of a triple-$\bfq$ skyrmion lattice~\cite{kurumaji2019skyrmion}. 

\subsection{DFT calculations}
\begin{figure}
    \centering
    \includegraphics[scale=0.32]{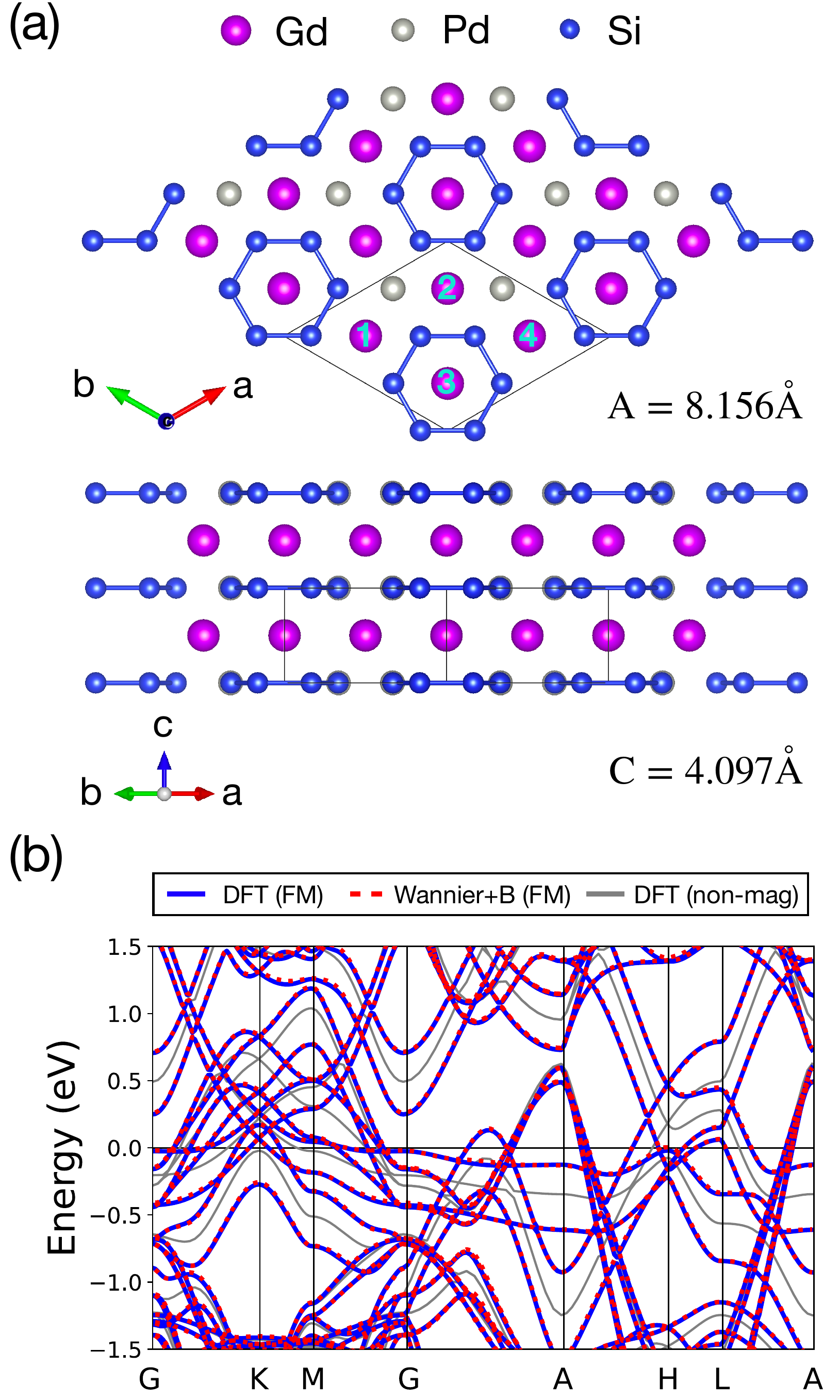}
    \caption{(a) Crystal structure and atomic position of $\rm Gd_2PdSi_3$ adopted in the calculation. In the top panel, we highlight the four Gd-atoms in a primitive unit cell (black line) on which the configuration of the localized spin determines the EX $B$-field. (b) Band dispersion along the high-symmetry path computed using different conditions and methods. Comparing the bands of the NM and FM phase, we can see that the EX $B$-field induces a $\sim 500$ meV spin-splitting. Combining the NM solution and the EX $B$-field, we can reproduce the FM band with a small error $<10\,$meV. }
    \label{Fig:crystal_bnd_fit}
\end{figure}

To acquire a microscopic understanding of this material, we apply the method introduced in Sec.~\ref{Sec:Theory} to compute the THE from first principles. In order to exclude the contribution of the conventional AHE, we use the scalar relativistic approximation for all our DFT calculations~\cite{yi2009skyrmions}. 
To reduce the computational cost, we employ the simplified crystal structure reported in a previous study \cite{hirschberger2020topological,spitz_structure_2022} that the unit cell contains two formula units of $\rm Gd_2PdSi_3$ with $\rm Pd$ and $\rm Si$ arranged in the same way as $\rm Co$ and $\rm Si$ in $\rm Ce_2CoSi_3$. 
The lattice constants are set to be $a=8.156 \AA$ and $c=4.097 \AA$ [Fig.~\ref{Fig:crystal_bnd_fit}(a), Ref.~\cite{hirschberger2020topological}]. 
\\
\indent
To apply the modeling approach presented in Sec.~\ref{Subsec:Model}, we assume  that the EX $B$-field in the unit cell is exclusively determined by the localized Gd 4$f$ spins. 
We use the ansatz, Eq.~(\ref{Eq:B_JS}), and obtain the cofficient by fitting 128 EX $B$-field potentials computed by self-consistent DFT calculations for randomly generated initial spin configurations.
We leave the discussion of fitting process and validation on the fitted function in Appendix~\ref{Append:fitting} where we conclude that the EX $B$-field can be uniquely determined by the configuration of the four Gd 4$f$ spins in the unit cell, and its numerical form can be obtained from the ansatz function $\stackrel{\leftrightarrow}{\bf J}(\bfx)$ without performing self-consistent calculations.
\\
\indent
\begin{figure}[t]
    \centering
    \includegraphics[scale=0.28]{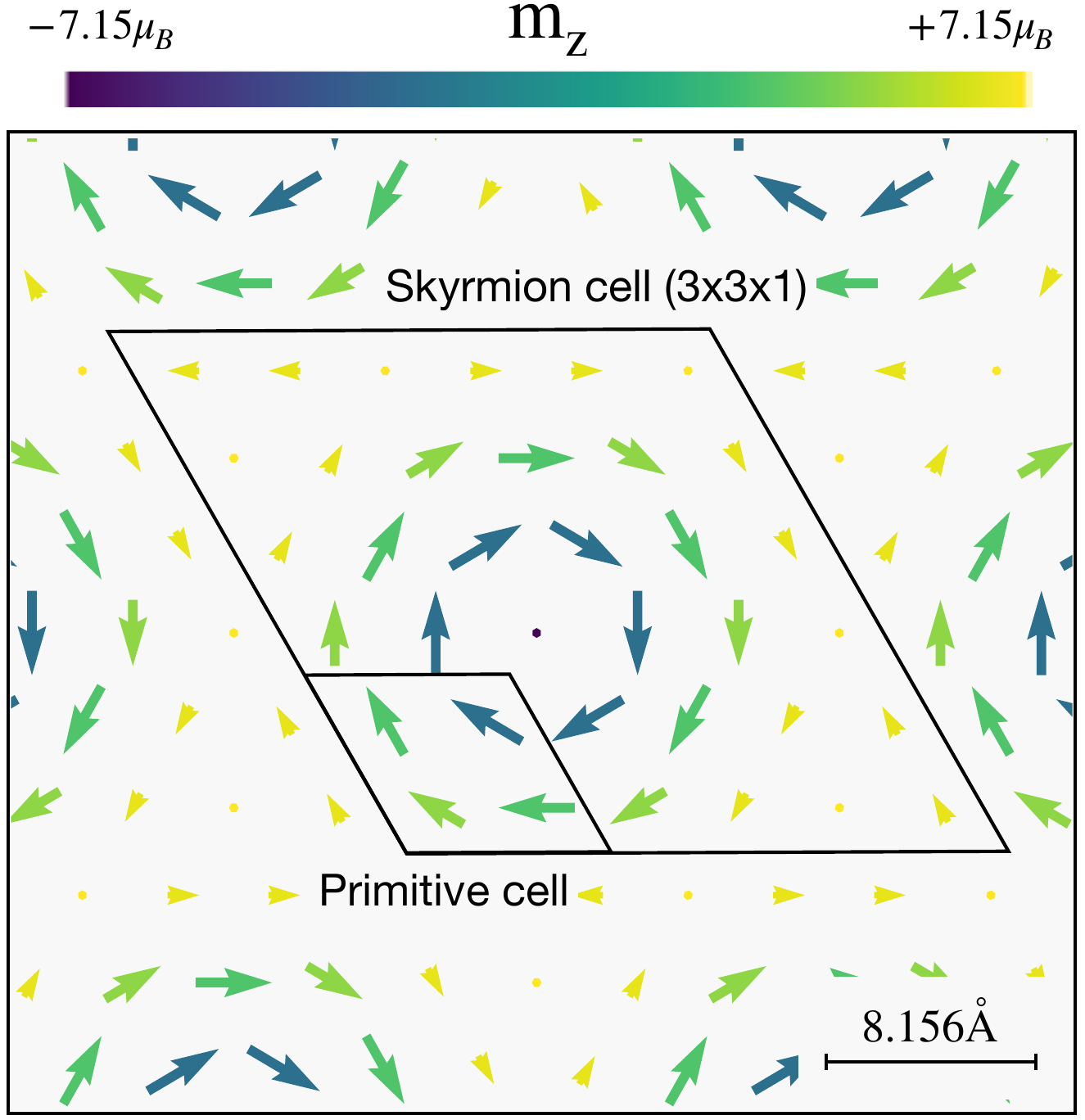}
    \caption{Skyrmion spin texture following Eq.~(\ref{Eq:sk-texture}). The arrows denote the magnetic moment of each Gd atom. The length of these arrows indicates the in-plane projection and the color map corresponds to the out-of-plane ($z-$)component. Due to this magnetic structure, the translation symmetry is partially broken, and a $3\times 3$ larger unit cell forms in skyrmion phase.}
    \label{Fig:sk_spin}
\end{figure}
 We next construct a Wannier TB model using the Bloch wave functions for the NM-phase. In order to describe the magnetic skyrmion or field-polarized phase, the NM-Wannier orbitals must form a complete set, at least for states near the Fermi level. To verify the completeness, we set all the spins in Eq.~(\ref{Eq:B_JS}) to the $z$-direction to obtain the FM EX $B$-field and add its effect to the NM-TB model via Eq.~(\ref{Eq:mBm}), which is expected to be equivalent to the FM-TB model constructed from the Bloch wave functions for the FM state. In Fig.~\ref{Fig:crystal_bnd_fit}(c), we plot the band structure computed from the FM-TB model and the one obtained by the self-consistent DFT calculation. The result shows a good agreement between the two methods, validating the reliability of our approach.
\\
\indent
To discuss the skyrmion phase, we construct the spin texture using \cite{kurumaji2019skyrmion}
\begin{equation}
    {\bf S}(\bfR)=M_0 {\bfe}_z+\sum_{i=1,2,3}(\bfm_i e^{i \bfQ_i\cdot \bfR}+c.c.)
    \label{Eq:sk-texture}
\end{equation}
with the magnetic modulation:
\begin{equation}
    \bfm_i=(m_z {\bf e}_z+i m_\perp  \hat{\bfQ}_i\times {\bf e}_z)
\end{equation}
where $\bfR$ and ${\bf e}_z$ are the position vector of the Gd 4$f$ localized spins and the unit vector along the $z$ direction, respectively. We set $m_z=m_\perp < 0$ and $M_0$ according to the experimental data and fix $|{\bf S}(\bfR)|=7.12~\mu_B$ according to the self-consistent DFT calculation. Since the experimental value of the skyrmion size ($\sim$2.5~nm) is about three times larger than the lattice constant of the primitive cell \cite{kurumaji2019skyrmion}, we assign the modulation wave vector by:
\begin{align}
    &\bfQ_1=\frac{1}{3}(-\sqrt{3}/2,-1/2,0)\nn\\
   & \bfQ_2=\frac{1}{3}(\sqrt{3}/2,-1/2,0)\nn\\
    &\bfQ_3=\frac{1}{3}(0,1,0)
\end{align}
such that the skyrmion lattice is approximated to be a commensurate superlattice of $3\times 3\times 1$ the primitive Gd$_2$PdSi$_3$ cell of the NM state [Fig.~(\ref{Fig:sk_spin})]. 
\\
\indent
\begin{figure}[t]
    \centering
    \includegraphics[scale=0.22]{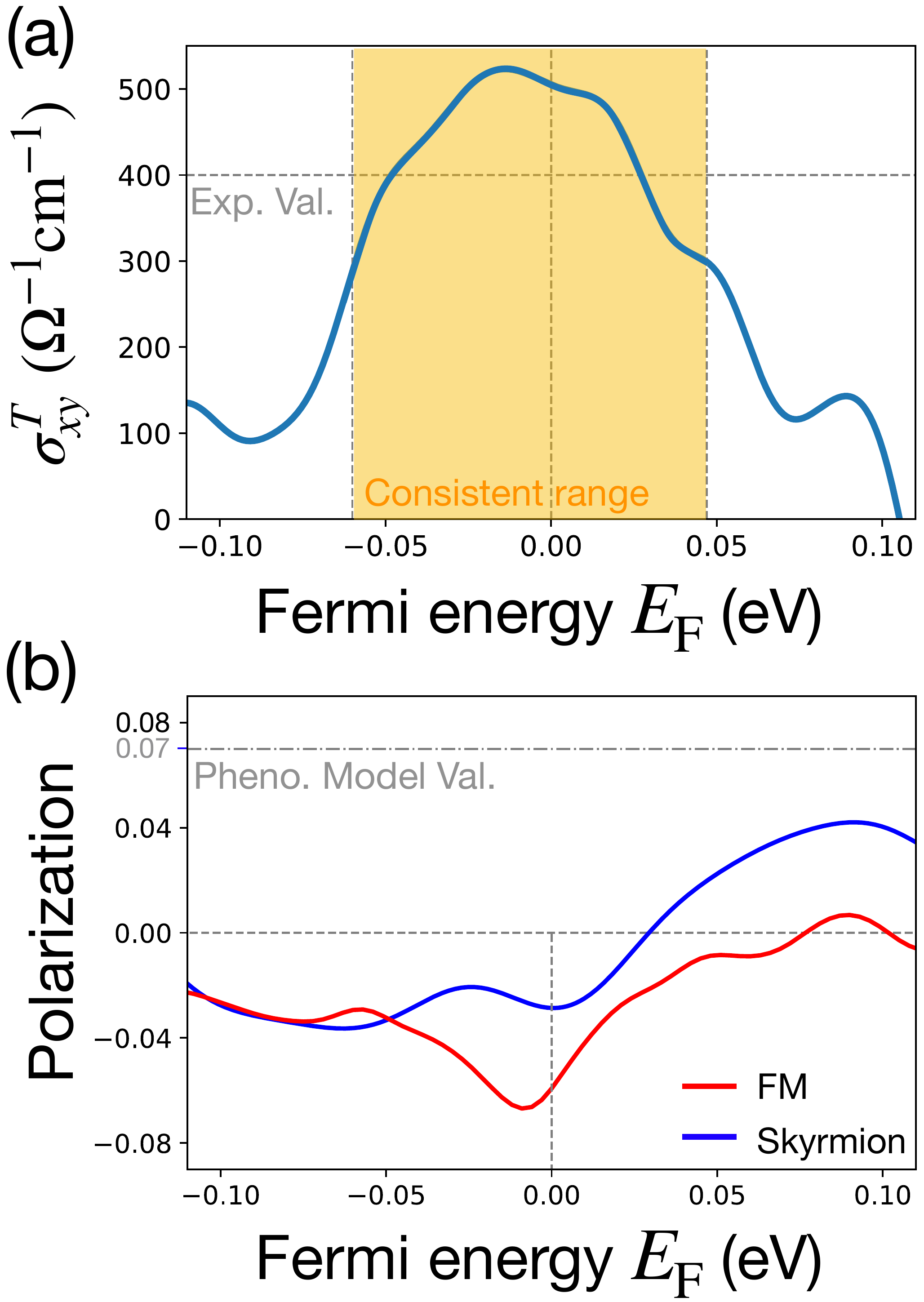}
    \caption{(a) The computed topological Hall effect (THE) as a function of  the Fermi energy $E_\mathrm{F}$. The zero-value of $E_\mathrm{F}$ is set in accordance with chemical analysis of sample used for Hall effect measurements (Appendix \ref{Append:EF_shift}). The computed $\sigma^T_{xy}=500~ \Omega^{-1}{\rm cm^{-1}}$ is broadly consistent with the experimental Hall conductivity measurement (yellow highlighted range). (b) The spin polarization $P$ in the skyrmion lattice and FM phases as a function of the band filling (Fermi) energy. Within a reasonable range of band filling energies, no value of $E_\mathrm{F}$ can reproduce the experimental spin polarization $P=+0.07$, which is obtained from the data based on a phenomenological model.}
    \label{Fig:sigma_scan_EF}
\end{figure}
Combining Eq.~(\ref{Eq:B_JS}) with Eq.~(\ref{Eq:sk-texture}), and setting the EX $B$-field into the TB Hamiltonian [Eq.~(\ref{Eq:HB})], we solve the skyrmion Hamiltonian and obtain the electronic structure of the skyrmion phase. 
From the result, we use the Kubo formula in the static limit, Eq.~(\ref{Eq:Kubo}), to calculate the Hall conductivity. Following the standard analysis procedure \cite{pizzi2020wannier90}, we scan the Hall conductivity as a function of the Fermi level [Fig.~\ref{Fig:sigma_scan_EF}(a)]. At the Fermi level $E_\mathrm{F}$ corresponding to the experimental condition (Appendix \ref{Append:EF_shift}), we obtain a THE of $\sigma^T_{xy}=500~ \Omega^{-1}{\rm cm^{-1}}$ which is consistent with the experimental value of $\sigma^{T}_{xy}=400~ \Omega^{-1}{\rm cm^{-1}}$ \cite{hirschberger2020topological}. 
\\
\indent
Let us finally discuss the validity of the phenomenological model conventionally used to discuss the THE in skyrmion materials~\cite{neubauer2009topological,ritz2013giant}. In this model, the Hall resistivity is expressed as:
\begin{align}
    \rho_{xy}&=
   \rho^{\rm N}_{xy}+
   \rho^{\rm A}_{xy}+
   \rho^{\rm T}_{xy}
   \nn\\
   &=R_0B+\mu_0R_{\rm s}M+PR_0B_{\rm em}
   \label{Eq:Hall_pheno}
\end{align}
where $\rho_{xy}^{\rm N}$, $\rho_{xy}^{\rm A}$, and $\rho_{xy}^{\rm T}$ denote the contributions to the resistivity due to the normal Hall effect (NHE), AHE, and THE, respectively. 
$P$ is the spin-polarization at the Fermi level and
$R_{0}$ and $R_{\rm s}$ are empirical parameters termed normal Hall coefficient and anomalous Hall coefficients, respectively. $B_{\rm em}$ is the emergent magnetic field from the skyrmion, which is proportional to the topological charge ($B_{\rm em}=4\pi n_{Sk}$, from Eq.~(\ref{Eq:Nsk})) \footnote{ The emergent $B_{em}$ raises from the noncoplaner precession which is different from the local exchange $B$-field, $B_{\rm KS}$, induced by local magnetization. }. The last term indicates that the THE exists only in systems with asymmetric up/down-spin occupation, $P\neq0$.
\\
\indent
We evaluate the spin polarization at the Fermi level ($E_{\rm F}$) from the eigenfunctions of the nonmagnetic TB model:
\begin{equation}
    {\bf P}(E_{\rm F})=
    \left.\sum_{n \bfk}\<n\bfk|\sigma_z|n\bfk\>
    \delta(E_{n\bfk}-E_{\rm F}) \middle/\sum_{n \bfk} \delta(E_{n\bfk}-E_{\rm F})\right. .
\end{equation}
In Fig.~\ref{Fig:sigma_scan_EF}(b), we show the spin polarization in the skyrmion and FM phase as a function of the Fermi level. In both cases, the results show that $P$ is smaller than $0.04$ over a wide range of energies and takes negative values at the experimentally determined $E_\mathrm{F}$. Although this number is of comparable magnitude to $P_\mathrm{exp} = +0.07$ deduced from experiments using the phenomenological model of Eq.~(\ref{Eq:Hall_pheno}), its sign and size nevertheless sharply conflict with Eq.~(\ref{Eq:Hall_pheno})~\cite{kurumaji2019skyrmion,hirschberger2020topological,neubauer2009topological}. 
This difference in the polarization indicates the failure of the phenomenological description of the THE in $\rm Gd_2PdSi_3$. 
\subsection{Discussion}
\begin{figure*}[t]
    \centering
    \includegraphics[scale=0.25]{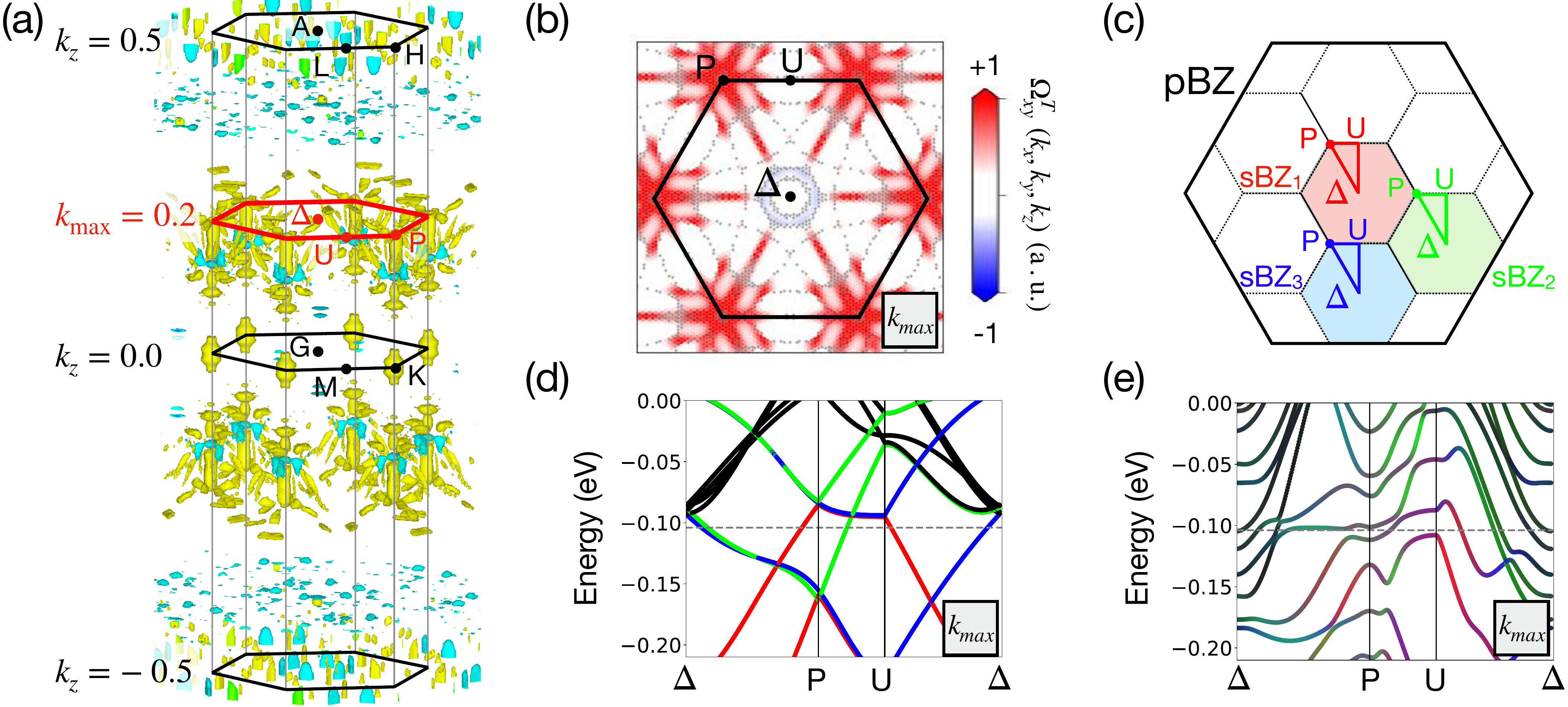}
    \caption{(a) Isosurfaces of contributions to the topological Hall effect (THE) in the three-dimensional Brillouin zone (BZ) of Gd$_2$PdSi$_3$. The positive part, colored in yellow, dominates and mainly resides along the K-H line for $k_z=0.1\sim 0.2$. In contrast, the negative part (colored in blue) has a minor contribution and is distributed around $k_z=0.4$. (b) Berry curvature contour for the cross-section at $k_z=k_{\rm max}$. Centered at the K-point, strong Berry curvature extends along the $\Delta$-P and P-U lines. (c) Illustration of BZ folding for a hexagonal system from a primitive cell (pBZ) to a $3 \times 3 \times 1 $ supercell (sBZ). Due to a three-fold rotation symmetry, eigenstates at the P point of the three sBZ's (colored red, green, and blue) are (three-fold) degenerate. On the other hand, states along the $\Delta$-P line in the green and blue sBZ are connected by the mirror symmetry (two-fold degeneracy). Similar relations apply for the states along the P-U lines for red and blue sBZ's. (d) Band structure of $\rm Gd_2PdSi_3$ along high-symmetry paths in the nonmagnetic (NM) pBZ, at $k_z=k_{\rm max}$. The colors denote the origin of bands from the sBZ's specified in (c), while the black bands are from sBZ's other than the three colored sBZ's. The degeneracy pattern follows the symmetry argument made in (c). (e) Band structure at $k_z=k_{\rm max}$, in the skyrmion phase, colored as in (d). The skyrmion potential breaks the band degeneracy and mixes the wave functions originating from different sBZ's. Significant color changes within each band from $\Delta$ to P and P to U indicate strong $\mathbf{k}$-dependent wave function mixing. As the Fermi level lies within the band gap at P, we observe a large Berry curvature, as shown in (b).  }
    \label{Fig:combine}
\end{figure*}
In order to understand the THE contributions in the Brillouin zone, we decompose $\sigma^{\rm T}_{xy}$ as a function of wave vector $\bfk$ and plot its isosurface in the three-dimensional hexagonal BZ in Fig.~\ref{Fig:combine}(a). We found a significant positive contribution at the K corner for $k_z$ from 0.1 to 0.2 with no concentrated negative contribution. Furthermore, to have a clear in-plane visualization, we select the cross-section at $k_z=k_{\rm max}$ with the largest THE contribution to plot the Berry curvature within the $k_x$-$k_y$ plane. As shown in Fig.~\ref{Fig:combine}(b), we can also observe significant contribution along the $\Delta$-P and P-U line \footnote{Besides the conventional notations for high symmetry points in the hexagonal BZ, we use $\Delta$ for points in the G-A interval, P for points in the K-H interval, and U for points in the M-L interval. See Fig.~\ref{Fig:combine}(a). }. 
\\
\indent
Taking the analogy with AHE, we conjecture that a nodal-line breaking mechanism occurs along high-symmetry lines. To verify the proposition, we plotted out the band folding boundary in Fig.~\ref{Fig:combine}(c) and highlighted that among the nine folded sBZ, the P points in three of them (colored red, green, and blue) are connected by three-fold rotation symmetry, resulting in a three-fold degeneracy at the P point in the sBZ. However, the rotation symmetry no longer applies when moving away from the P point along the P-$\Delta$ line. Instead, the mirror symmetry maps the P-$\Delta$ lines of the blue and green sBZ into each other, thus maintaining a two-fold degeneracy. A similar argument applies to the P-U line, where the degeneracy arises from the blue and red sBZs. {To visualize these degenerating behaviors, we take the cross-section $k_z=k_{\rm max}$ as an example and plot the band structure along the $\Delta$-P-U-$\Delta$ path in the NM phase in Fig.~\ref{Fig:combine}(d). We colored the bands with red, green, and blue if they are derived from the sBZ specified in Fig.~\ref{Fig:combine}(c). The overlapping of the green and blue bands along the $\Delta$-P path and the overlapping of the red and blue bands along the P-U path are consistent with our symmetry argument for the degeneracy.} 
\\
\indent
These degenerate points (lines) are lifted when the skyrmion potential is introduced, breaking the symmetry. 
{The resultant skyrmion band structure is presented in Fig.~\ref{Fig:combine}(e).} 
Compared with Fig.~\ref{Fig:combine}(d), we can observe that the degeneracy at the P point is now splitting into multi-band gaps, while the green and blue bands mix into blue-green lines, and the red and blue bands become purple lines.
Further, by comparing Fig.~\ref{Fig:combine}(e) with Fig.~\ref{Fig:combine}(b), we can identify the large Berry curvature region to the intersection between Fermi level and the band gaps at the P point. Based on the rapid color change of bands on two sides of the Fermi level, we can attribute the large THE to wavefunction mixing among states belonging to different sBZs. It is worth noting that outside the range of $k_z=0.1\sim0.2$, the degeneracy moves away from the Fermi level, so the gap opening no longer contributes to the THE. 
\\
\indent
Overall, based on our first-principles calculation along with subsequent analysis, we identify the THE in $\rm Gd_2PdSi_3$ as a result of nodal line breaking along the high-symmetry lines induced by the EX $B$-field generated by the topological chiral spin textures.
\\
\indent
\section{conclusions \label{sect:conclusions}}
\vspace{-10pt}
In this work, we present a systematic approach to compute the topological Hall effect (THE) from first principles, based on density functional theory. In the first part, we develop a modeling method to overcome the scaling problem of a large supercell, which enables us to obtain the EX $B$-field of any spin orientation from minimal supercell calculations. The model relates the localized spin and local EX $B$-field in a real space grid, without requiring its definite form; this makes our approach general and scalable. We can also expect that machine learning methods such as neural networks or unsupervised learning could enhance accuracy and reduce the need for extensive data generation~\cite{fiedler2022deep}. In the second part, we introduce a mapping scheme to project the unit cell Wannier TB model onto a supercell system, in which we can incorporate the EX $B$-field. The resultant Hamiltonian provides all information of the electronic structure including eigenenergies and wave functions, from which we can calculate physical properties such as the Hall conductivity.
\\
\indent
After setting up the formalism, we apply the scheme to the calculation of the THE in the skyrmion phase of $\rm Gd_2PdSi_3$. Our results challenge the validity of the conventional phenomenological model of the THE, as there is a noticeable discrepancy between the spin polarization $P$ at the Fermi level, calculated by our method, and the value of $P$ derived from phenomenological approaches. On the other hand, utilizing the Kubo formula, we calculate a THE of $\sigma_{xy}\sim 500~ \Omega^{-1}{\rm cm }^{-1}$, in good agreement with the experiment value. To provide a microscopic insight into its origin, we analyze the Berry curvature in the BZ and reveal that a large contribution comes from the crossing points along high-symmetry paths between back-folded bands. 
\\
\indent
Overall, our work establishes a framework for predicting the topological Hall effect (THE) induced by topologically nontrivial spin textures. The present calculations serve as a benchmark for distinguishing the THE from the AHE, as well as from extrinsic effects, and assist the identification of the skyrmion formation in larger classes of materials. It should be noted that the present formalism is not restricted to skyrmions but applicable to any spatially extended spin systems. For future development, we aim to generalize the formalism to study magnetism characterized by $d$-electrons, such as B20-MnSi or -FeGe. 
\begin{acknowledgments}
\vspace{-10pt}
We are grateful for fruitful discussions with YiZhou Liu. H.-Y. Chen is supported by Riken Special Postdoctoral Researchers Program. This work was supported by the RIKEN TRIP initiative (RIKEN Quantum, AGIS, Multi-Electron Group).
We also acknowledge the financial support by Grant-in-Aids for Scientific Research (JSPS KAKENHI) Grant Numbers JP21H04990, JP22F22742, JP22H00290, JP23H05431, JP24H01607, and JP24K00581 as well as JST-CREST Nos.~JPMJCR23O4, JPMJCR1874, and JPMJCR20T1; JST-ASPIRE No.~JPMJAP2317; JST-Mirai JPMJMI20A1; and JST FOREST Grant No. JPMJFR2238.
\end{acknowledgments}

\appendix
\section{Implementation detail for Sec. \ref{Subsec:Model} \label{Append:implement}}
\vspace{-10pt}
Here, we provide detailed information about how to extract the necessary quantities for implementing the theoretical approach depicted in Sec.~\ref{Subsec:Model} using the {\sc QUANTUM ESPRESSO} package \cite{giannozzi2009quantum}.  For generating data to fit Eq.~(\ref{Eq:B_JS}), we assign random values to the "{\it starting\_magnetization}" flag in the input for the {\it pw.x} executable. After the scf calculations are finished, Eq.~(\ref{Eq:S-m}) is automatically computed, and the numerical values of $\bfS^{a,\d}_r$ are printed in the standard output. On the other hand, the self-consistent EX $B$-field will not be saved; instead, it is computed by the "{\it set\_vrs}" subroutine whenever needed. Therefore, we modify the code "{\rm set\_vsr.f90}" to force the "{\it set\_vrs}" subroutine to print out its output "{\it vrs}", for which the first component is the EX potential $V_{\rm KS}(\bfx)$ and the second to fourth component give the EX $B$-field, ${\bfB}_{\rm KS}(\bfx)$.

\section{Numerical method \label{Append:numerical}}
\vspace{-10pt}
We carry out DFT calculations by using the {\sc QUANTUM ESPRESSO} package \cite{giannozzi2009quantum} with scalar relativistic norm-conserving pseudopotential under generalized gradient approximation (GGA) generated with Pseudo-dojo \cite{perdew1996generalized,perdew2008restoring,van2018pseudodojo}. We employ a $4\times 4\times 8$ k-grid for BZ sampling in self-consistent field (SCF) calculations with 110 Ry energy cut-off which corresponds to a $48\times 48\times 96$ real space grid after the Fourier transformation. The Wannier-TB model is constructed based on a non-self-consistent (NSCF) calculation on a $4\times 4\times 6$ k-grid using $140$ Wannier orbitals in the NM phase using the {\sc Wannier90} package \cite{pizzi2020wannier90}. We use a $\pm 15.0$ eV window centering at the Fermi level for the band disentanglement. For the calculation of the Hall conductivity, we use the Kubo formula with a regular sampling on a $192\times 192\times 288$ interpolated \textbf{k}-grid and with a broadening parameter $\delta=20$ meV.
{
\section{Details for fitting $\stackrel{\leftrightarrow}{\bf J}$-function \label{Append:fitting}}
As pointed out in the main text, for $\rm Gd_2PdSi_3$, we assume $V_{\rm KS}$ is independent of the spin orientation and $\bfB_{\rm KS}$ can be uniquely obtained by the four spins on the Gd-atoms within a primitive unit cell. Using the tensor function in Eq.~(\ref{Eq:B_JS}), we take a minimal ansatz with $a=1-4$, $\delta=0$, so there are 12 parameters to determine the $B$-field in each direction. We generated 128 data sets with random spin configuration to fit the function. Within the 128 potentials, we found that the variation in $V_{\rm KS}$'s is less than $10^{-5}$, which is consistent with the assumption made for Eq.~(\ref{Eq:KS-noncol-decomp}). This result is expected for the Gd-based material with half-filled 4$f$-electrons. However, such a decomposition of scalar potential from the spins may not be valid for other magnets like Nd- or Eu-based compounds, for which an ansatz for the scalar potential is also required. 
\begin{figure}[h]
    \centering
    \includegraphics[scale=0.3]{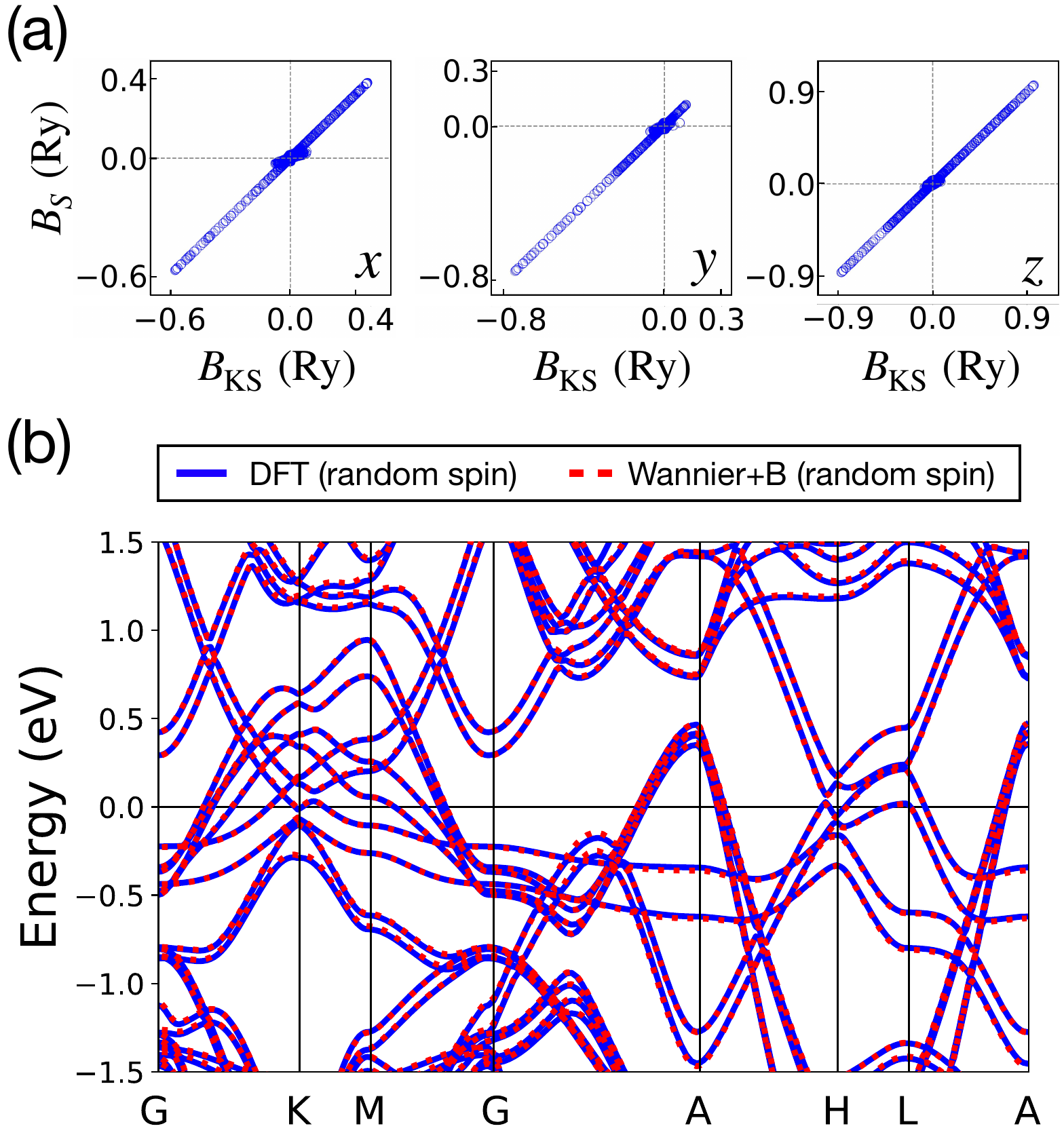}
    \caption{{(a) We contrast the exchange magnetic fields (EX $B$-fields) for a randomly chosen spin configuration given in Eq.~(\ref{Eq:S_random}), computed by self-consistent DFT ($B_\mathrm{KS}$) with the one generated from our fitted function, $B_S$ in Eq.~(\ref{Eq:B_JS}). The two results agree well with each other, except for a minor error in the weak field region. (b) The band structure of $\rm Gd_2PdSi_3$  computed by DFT (blue line) and from a TB model plus the fitted EX $B$-field (red dashed line).}}
    \label{Fig:fit-random}
\end{figure}
On the other hand, for the self-consistent localized spin on Gd, we found a variation $\Delta |\bfS(\bfR)| \approx 0.015$ $\mu_B$ which contributes an error less than $10^{-3}$ when we fix $|\bfS(\bfR)|$ at its average value 7.12 $\mu_B$. It is worth noting that the spin orientation does not change much compared to the initial value assigned by the "{\it starting\_magnetization}"-flag, even though we did not add any penalty energy to constrain the orientation during the SCF computation.
\\
\indent
In Fig.~\ref{Fig:crystal_bnd_fit}(b), we have shown that the fitted function for the EX $B$-field can accurately reproduce the DFT band structure in the FM phase. For a general spin configuration, we verified the fitting function, $\stackrel{\leftrightarrow}{\bf J}(\bfx)$, by comparing the model $B$-field with the one evaluated by self-consistent DFT calculations, using a randomly chosen spin set:
\begin{align}
    &\bfS^{1,2}=( -0.20 ,  -0.29   ,-7.14)&\nn\\
    &\bfS^{1,2}=(  -2.83  , ~ ~5.74   ,~ ~3.16)&\nn\\
    &\bfS^{1,3}=(   -2.54  , ~ ~1.80  , ~~ 6.43)&\nn\\
    &\bfS^{1,4}=(  ~~ 4.24 ,  -0.85 ,  -5.70)&
    \label{Eq:S_random}
\end{align}
and present the result in Fig.~\ref{Fig:fit-random}(a). We found the two functions, $\bfB_S$ and $\bfB_{\rm KS}$, to be almost identical except for the region of small $B$-fields, which does not make significant contributions to the Berry curvature or the Hall conductivity. Furthermore, we accommodate the fitted $B$-field into the Wannier TB model to compute the band structure and plot it in Fig.~\ref{Fig:fit-random}(b) along with the bands obtained by direct DFT calculations. Compared to the FM case (Fig.~\ref{Fig:crystal_bnd_fit}(b)), we can see that the spin-splitting is reduced due to the noncolinear spin orientation; this reduction can also be correctly captured by the TB model. The observed agreement validates our modeling approach proposed in Sec. \ref{Subsec:Model}, even for arbitrarily oriented spin configurations.}

\section{Chemical composition and Fermi energy of Gd$_2$PdSi$_3$\label{Append:EF_shift}}
\vspace{-10pt}
\begin{figure}[b]
    \centering
    \includegraphics[scale=0.195]{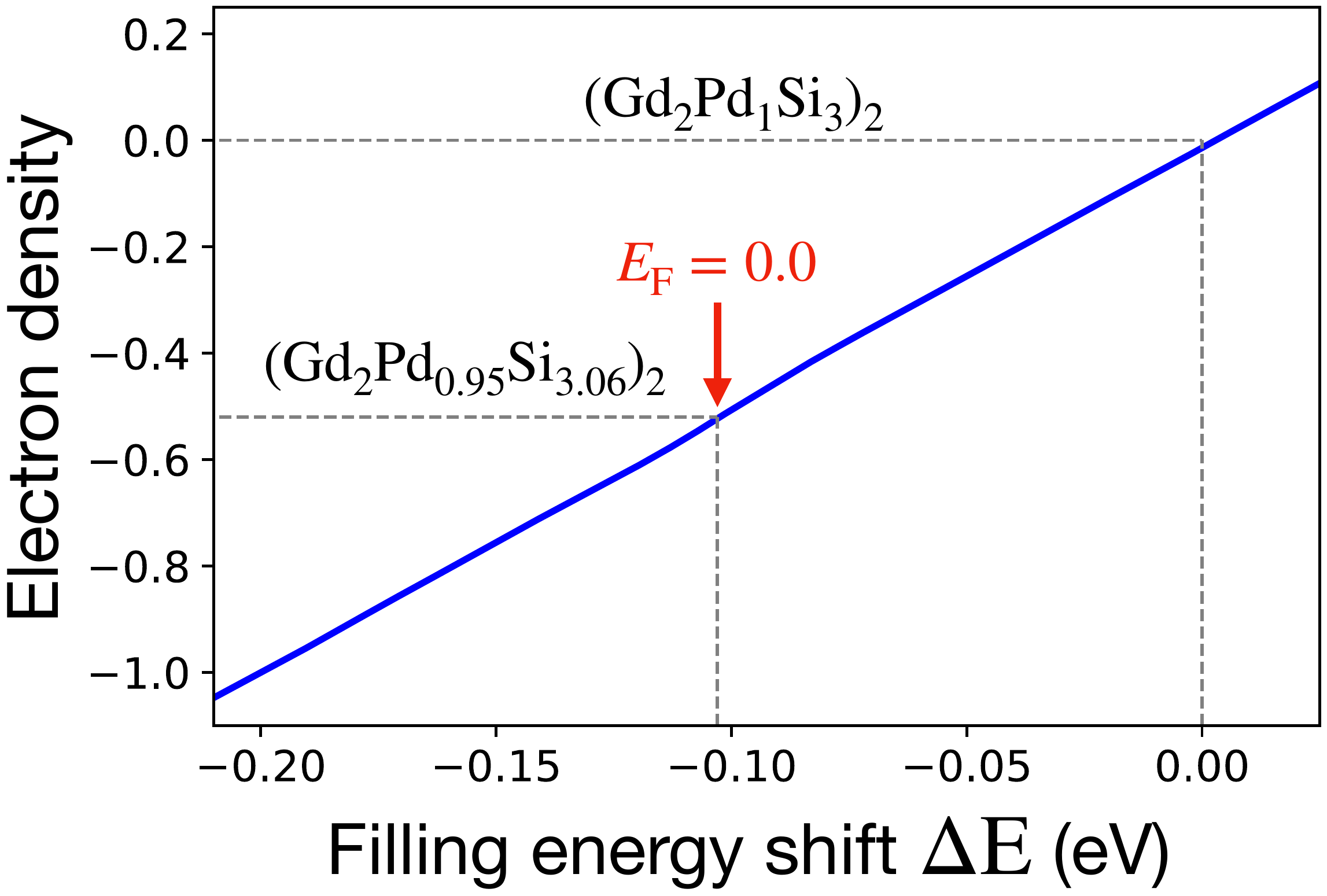}
    \caption{(Blue) Electron density as a function of band filling  energy. The baseline is set by the stoichiometric crystal of $\rm (Gd_2PdSi_3)_2$. Compared to the baseline, $\rm (Gd_2Pd_{0.95}Si_{3.06})_2$ has a depletion of $0.52$ electrons per chemical formula. Therefore, we set the filling energy $\Delta E = -104$ meV as the zero-point of the Fermi level throughout this work. }
    \label{Fig:scan_EF}
\end{figure}
The computed topological Hall conductivity $\sigma_{xy}^T$ shows a strong dependence on the Fermi-level $E_\mathrm{F}$ (see Fig.~\ref{Fig:sigma_scan_EF}), so that it is essential to determine the $E_F$ which precisely reflects the electronic occupation and stoichiometry of our sample. For Gd, Pd, and Si atoms, we have electronic structure:
\begin{align}
    {\rm Gd: [Xe]}& ~4f^7 ~5d^1 ~6s^2\nn\\
    {\rm Pd: [Kr]}& ~4d^{10}\nn\\
    {\rm Si: [Ne]}& ~3s^2~ 3p^2\nn
\end{align}
such that for each chemical formula $({\rm Gd_{2+x}Pd_{1+y}Si_{3+z}})_2$, the number of {valence} electrons is:
\begin{equation}
    \Delta n=(10x+10y+4z)\times 2.
\end{equation}

We estimate the actual valence electron density in our single crystal, which shows $\sigma_{xy}^T=400\,$S/cm~\cite{hirschberger2020topological}, based on {measurements of the chemical composition by SEM-EDX (energy-dispersive x-ray spectroscopy)}. {Indeed, EDX has a high uncertainty for lighter atoms, such as silicon, but is more precise for heavier atoms such as Pd and Gd; reasonable error bars are assumed for the following analysis.} {Based on EDX, a sample with $\sigma_{xy}^T=400\,$S/cm has the composition $x = 0$, $y = -0.05\,(2)$, $z=0.06\,(20)$ 
leading to a depletion of $0.5\,(2)$ electrons per primitive unit cell of the NM state. On the other hand, from the band structure, we can determine the Fermi level as a function of {valence electron density}. We present the relation in Fig.~\ref{Fig:scan_EF} and conclude that $E_\mathrm{F}$ is shifted by $-100\,(40)\,$meV as compared to the stoichiometric sample, due to slight palladium deficiency. In the calculation, we assume a Gd:Si ratio of 1:3, which is within the experimental error. We note that moderate, yet finite disorder is essential for suppressing extrinsic contributions to the Hall effect, and for making the intrinsic (Berry curvature) contribution visible~\cite{nagaosa2010anomalous}. Indeed, the conductivity $\sigma_{xx}$ of the single crystal presently used for Hall effect measurements is $\approx 14000\,$S/cm, well within the intrinsic regime identified in Ref.~\cite{nagaosa2010anomalous}.
\bibliography{THE_ref}
\end{document}